# Solar System evolution from compositional mapping of the asteroid belt

**Authors:** F. E. DeMeo[1,2,3*], B. Carry[4,5]

[1]Harvard-Smithsonian Center for Astrophysics, 60 Garden Street, MS-16, Cambridge, MA, 02138, USA.

fdemeo@cfa.harvard.edu

[2]Department of Earth, Atmospheric and Planetary Sciences, MIT, 77 Massachusetts Avenue, Cambridge, MA, 02139, USA.

[3]Hubble Fellow

[4]Institut de Mécanique Céleste et de Calcul des Éphémérides, Observatoire de Paris, UMR8028 CNRS, 77 av. Denfert-Rochereau 75014 Paris, France.

[5]European Space Astronomy Centre, ESA, P.O. Box 78, 28691 Villanueva de la Cañada, Madrid, Spain.

Advances in the discovery and characterization of asteroids over the past decade have revealed an unanticipated underlying structure that points to a dramatic early history of the inner Solar System. The asteroids in the main asteroid belt have been discovered to be more compositionally diverse with size and distance from the Sun than had previously been known. This implies substantial mixing through processes such as planetary migration and the subsequent dynamical processes.

Although studies of exoplanetary systems have the advantage of numbers[1] to answer the question of how planetary systems are built, our solar system has the advantage of detail. For nearly two centuries since their first discovery, asteroids have been viewed as remnants of planetary formation. Located between Mars and Jupiter in the main asteroid belt (Fig 1), it was perceived they formed essentially where they are now[2].

Early measurements showed asteroids in the inner part of the asteroid belt were more reflective and "redder" than the outer, "bluer" ones[4-6]. In the 1980s, distinct color groupings of major asteroid compositional types were discovered as a function of distance from the sun[2]. In the classic theory, this was interpreted as the remnant of a thermal gradient across the belt at the time the Solar System formed[2,7-9]. Understanding that gradient promised to hold clues to the initial conditions during planet formation.

Yet, over the course of the discovery of over half a million asteroids since the 1980s, the idea of a static Solar System history has dramatically shifted to one of great dynamic change and mixing. Driving this view, was the effect on the main belt of planetary migration models that aimed to recreate the structure of the rest of the Solar System, such as the orbits of the giant planets, Pluto and the Transneptunian objects, and the Jupiter Trojans (which reside in the L4 and L5 Lagrange points of Jupiter's orbit)[10-13].

As the planetary migration models evolved, so also new compositional characteristics of the main belt were uncovered through observation that were increasingly inconsistent with the classic theory. At first, just a few rogue asteroids were found to be contaminating the distinct groupings[14-16]. Now, with tens of thousands of asteroids to analyse for which we have compositional measurements[17,18], we see that this mixing of asteroid types is more of the rule, rather than the exception, across the belt[19].

Today, all the newly revealed aspects of the asteroid belt, including its orbital and compositional structure and the dynamical processes that sculpt it, contribute to a more coherent story. In modern dynamical models, the giant planets are thought to have migrated over substantial distances, shaking up asteroids that formed throughout the Solar System - like flakes in a snow globe, and transporting some to their current locations in the asteroid belt (Fig 2). The main asteroid belt thus samples the conditions across the entire Solar System. Yet at the same time the Hildas (located 4 AU from the Sun between the main belt and Jupiter (one astronomical unit is approximately the Earth– Sun distance); see Fig. 1) and Jupiter Trojans appear distinctly homogeneous, challenging us to untangle the various events of the Solar System's evolution. Our Solar System's path to creating the arrangement of the planets today and the conditions that made life on Earth possible will set the context for understanding the myriad of exoplanetary systems.

**Send in the Rogues**

Their generally redder-to-bluer color and compositional trend implied that asteroids tend to preserve their initial formation environment: the temperature and compositional gradient in that part of the disk at the time of planetesimal formation[2,9]. From what we understood at the time (the 1980s), guided by comparison with meteorites, the reddish (with a positive slope from ultraviolet-to-visible wavelengths) ones filling the inner main belt were melted igneous bodies[20], and the bluish (neutral slope from the ultraviolet-to-visible) ones in the outer main belt had undergone little thermal alteration[Error! Reference source not found.]. The goal of the next decade (the 1990s) was to explain how the thermal gradient could be so steep, creating such wildly different outcomes, from melted to primitive over a distance of just 1 au[21].

That original interpretation of the compositions of reddish and bluish asteroids was wrong. In fact, direct sampling (by spacecraft[22]) of the reddish asteroid (25143) Itokawa definitively showed that it did experience some heating but was relatively primitive compared with the previous interpretation of a melted body[23-27]. Though it was still a challenge to explain the asteroids' compositional and thermal trend from warm to cold, it was not as drastic a gradient as had been supposed.

Such compositional measurements for the largest asteroids seemed to better explain the gradient, but the few measurements becoming available for smaller objects were beginning to reveal the misfits. First was (1459) Magnya, a basaltic fragment discovered among the cold, bluish bodies[14]. Then, a handful more of these rogue igneous asteroids were found dispersed across the main belt[16,28,29]. Iron asteroids present in the main belt should have formed much closer to the Sun[15]. Primitive asteroids were discovered in the inner belt[30], and furthermore, the reddish objects extended out through the outer belt[31-33]. Other asteroids that appeared to be dry asteroids were discovered to contain volatiles on or just below the surface, suggesting that they formed beyond the snow-line (the distance from the Sun at which the temperature is low enough for water to be ice)[34-38]. At first, these observations seemed to represent 'contamination' by individual, unusual asteroids, but gradually it has become clear that even the core groups of reddish and bluish asteroids were more broadly distributed, further challenging the classic theory of a static Solar System.

**The compositional medley of asteroids**

Equipped with an abundance of visible-wavelength colors and surface brightness measurements from recent surveys[17,18] we can now reveal a new map for the distribution of asteroids down to diameters of 5km[19] (Fig 3). Traditionally, the distribution has been presented as the relative fraction of asteroid classes as a function of distance[2,9,31,32]. Now we compare bodies ranging from 5km to 1000km, so equally weighting each would distort the view. By transforming the map of the asteroid belt to the distribution of mass[19,39], we are able to accurately account for each asteroid type accurately, rather than the frequency or number of types (Fig 3). Furthermore, we can now explore the change in distribution as a function of size (Fig 4).

This is what we have found. The rarer asteroid types, such as the crust and mantle remnants of fully heated and melted bodies, are seen in all regions of the main belt[14,16]. We do not yet know if this means the locations of their respective parent bodies were ubiquitous in the inner Solar System or if they were created close to the Sun and later injected into the belt.[15,40]

Asteroids that look compositionally Trojan-like (D-types, see Fig. 3) are detected in the inner belt, where they are not predicted by dynamical models[19,41]. Their presence so close to the sun demands an explanation for how they arrived there and if they are even really linked to the Trojan asteroids at all.

The Hungaria region is typically associated with its eponymous and brightest member, (434) Hungaria, and similarly super-reflective asteroids[2,8] (E-types, see Fig. 3). Despite this, most of the mass of this region is contained within a few reddish and bluish objects, which are also common elsewhere in the main belt[42-44].

The relative mass contribution of each asteroid class changes as a function of size in each region of the main belt. Most dramatic is the increase of bluish objects (C-types, see Fig. 3) as size decreases in the inner belt. Although these bluish objects are notoriously rare in the inner belt at large sizes[2,32] where they comprise only 6% of the total mass, half of the mass is bluish at the smallest sizes.

In the outer belt, reddish asteroids (S-types, see Fig. 3) makeup a small fraction of the total there, yet the actual mass is still quite significant. In fact, we now find more than half of the mass of reddish objects outside of the inner belt[19].

Just over a decade ago, astronomers still clung to the concept of an orderly compositional gradient across the asteroid belt[45]. Since that time, the trickle of asteroids discovered in unexpected locations has turned into a river. We now see that all asteroid types exist in every region of the main belt (see Box 1 for discussion of Hildas and Trojans). The smorgasbord of compositional types of small bodies throughout the main belt contrasts with compositional groupings at large sizes. All these features demanded major changes in the interpretation of the history of the current asteroid belt and, in turn, of the Solar System.

**Cracking the 'compositional code' of the map**

Earlier planetesimal-formation theories that explained the history of the asteroid belt invoked turbulence in the nebula, radial decay of material due to gas drag, sweeping resonances, and scattered embryos[46,47]. Individually, each mechanism was, however, insufficient, and even together, although many of these mechanisms could deplete, excite, and partially mix the belt, they could not adequately reproduce the current asteroid belt[48].

The concept of planetary migration - whereby the planets change orbits over time owing to gravitational effects from the surrounding dust, gas, or planetesimals - was not new, but its introduction as a major driver of the history of the asteroid belt came only recently. Migration models began by explaining the orbital structure and mass distribution of the outer Solar System including the Kuiper Belt past Neptune[49]. Individual models could successfully recreate specific parts, but we still sought to define a consistent set of events that would explain all aspects of the outer Solar System. Every action of the planets causes a reaction in the asteroid belt, so these models also needed to be consistent with the compositional framework within the main belt that we see today.

The Nice Model was the first comprehensive solution that could simultaneously explain many unique structural properties of the Solar System[11-13,50,51] such as the locations of the giant planets and their orbital eccentricities[11], capture of the irregular satellites of Saturn[52], and the orbital properties of the Trojans[12] (Fig 2). In the original model, Jupiter moves inward while the other giant planets migrate outward. As Jupiter and Saturn crossed their 1:2 mean motion resonance, the system is destabilized[11]. In the most recent version of this model, the interaction between the giant planets and a massive, distant Kuiper disk causes the system to destabilize[13]. At that point, the primordial Jupiter Trojan region was emptied. Bodies that were scattered inward from beyond Neptune then repopulated this region. By reproducing the Trojans' orbital distributions and mass, the Nice model also naturally explains the why the Trojan region is compositionally distinct from the main belt: it would be populated solely by outer Solar System bodies and would not contain locally formed asteroids.

Missing from the Nice model, however, was an explanation of the large-scale mixing of reddish and bluish material in the asteroid belt that was becoming increasingly prominent. The Grand Tack model[53] showed that during the time of terrestrial planet formation (before the Nice model would have taken place), Jupiter could have migrated as close to the Sun as Mars is today. Jupiter would have moved right through the primordial asteroid belt, emptying it and then repopulating it with scrambled material from both the inner and outer Solar System as Jupiter reversed course and headed back toward the outer Solar System. Once the details of the resulting distribution of the Grand Tack Model have been closely compared to the emerging observational picture, it will become clear whether this model can crack the asteroid belt's compositional code.

Planetary migration ends well within the first billion years of our Solar System's 4.5 billion year history. The asteroid belt, however, is still dynamic today. Collisions between asteroids are continuously grinding down the bodies to smaller and smaller sizes. The smaller ones (< 40km) are then subject to the Yarkovsky effect, according to which uneven diurnal heating and cooling of the body alters the orbit[54-58]. The Yarkovsky effect thoroughly mixes small bodies within each section of the belt, but once they reach a major resonance—such as the 3:1 and 5:2 mean motion resonances at the locations where the orbital periods of an asteroid and of Jupiter are related by integers—they are swiftly ejected from the main belt[56-58]. Current observations[59,60] and models[28,61-63] indicate the strong resonances with Jupiter inhibit crossing of material from one region to another. These processes continue to mould the asteroid belt, erasing some of its past history and creating new structures in this complex system.

New observational evidence that reveals a greater mixing of bodies supports the idea of a Solar System that was and continues to be in a state of evolution and flux. Indeed, dynamical models have been leading us step-by-step to interpret the asteroid belt as a melting pot of bodies arriving from diverse backgrounds. Dynamical models have come a long way, but they have yet to explain the dichotomy between the orderly trend among the largest asteroids and the increased mixing of asteroid types at smaller sizes. Particularly noticeable are the scatter of igneous bodies and the existence of asteroids that look physically similar to Trojans in the inner belt. These details promise to teach us how our solar system was built, providing context for other planetary systems.

**The future**

The ultimate goal of asteroid studies is to complete the picture of where these bodies formed and how they relate to the current chemistry and volatile abundance for Earth. No longer is our Solar System just an isolated example, and with only a minimal speculative extrapolation, asteroid-like building blocks seem likely to have influenced countless terrestrial-like planetary systems. The ongoing hunt for Earth-like planets has as its corollary the hunt for possible signatures of asteroid-like zones and an assessment of their uniqueness or commonality in all planetary systems.

Even though we now know asteroid distributions in the Solar System down to 5km, we are literally only scratching the surface of what we can know about them. Asteroid interiors are the *terra incognita* for the next generation of asteroid researchers. At present we are frustrated by the inability of most physical measurements to provide any information on the interior of an asteroid. An asteroid's interior reveals its thermal history, which constrains the initial conditions of the protoplanetary disk during planetesimal formation. NASA's Dawn spacecraft mission recently provided a glimpse inside Vesta measuring its core mass fraction from shape and gravity measurements[64]. When Dawn visits Ceres, we will learn to what extent it differentiated into an ice mantle and rocky core[65-67]. Increased measurements of asteroid densities, provided mainly by the study of binary asteroids, will help us to infer their interior structure[39].

Each of our broad asteroid classes probably encompasses a wide variety of surface compositions[68,69]. Our meteorite collection has provided significant detail about the range of asteroid compositions, but to make firm links between the asteroids and meteorites, we need to observe asteroids in space and then measure that same body in a laboratory. This will be achieved by asteroid sample return missions that are already underway[70-72], as well as ``free sample return'' by meteorite falls such as the serendipitously discovered

Almahata Sitta meteorite (formerly asteroid 2008 TC$_3$ )[73,74].

Finally, the next step in distribution trends is to complement a refined understanding of asteroid compositions with physical measurements capable of detecting that detail on a large scale. The compositional trends discussed up to now cover broad taxonomic classes and combine objects into a small just a few major groups that do not accurately reflect the complexity of the asteroids' original and current compositions. Higher-spectral-resolution large-scale surveys at visible[75,76] and near- to mid-infrared wavelengths combined with already available albedo information for hundreds of thousands of asteroids would be the most realistic dataset to attain over the next decade or two.

**Acknowledgements**

We are grateful to R. Binzel for help shaping this review, and to K. Walsh, W. Bottke, N. Moskovitz, D. Polishook, T. Burbine, J. Wisdom, and A. Morales for discussions. We thank referee Clark Chapman. We acknowledge support from the ESAC faculty for FED's visit. This material is based upon work supported by the National Science Foundation under Grant 0907766 and by the National Aeronautics and Space Administration under Grant No. NNX12AL26G. Any opinions, findings, and conclusions or recommendations expressed in this material are those of the authors and do not necessarily reflect the views of the National Science Foundation or the National Aeronautics and Space Administration. Support for this work was provided by NASA through Hubble Fellowship grant HST-HF-51319.01-A awarded by the Space Telescope Science Institute, which is operated by the Association of Universities for Research in Astronomy, Inc., for NASA, under contract NAS 5-26555.

**Contributions**

Both authors worked jointly on the scientific analysis that resulted in Figs. 3 and 4. FED led the manuscript writing effort while BC created the figures.

Reprints and permissions information is available at www.nature.com/reprints

The authors declare no competing financial interests.

Correspondence and requests for materials should be addressed to fdemeo@cfa.harvard.edu

**Figures**

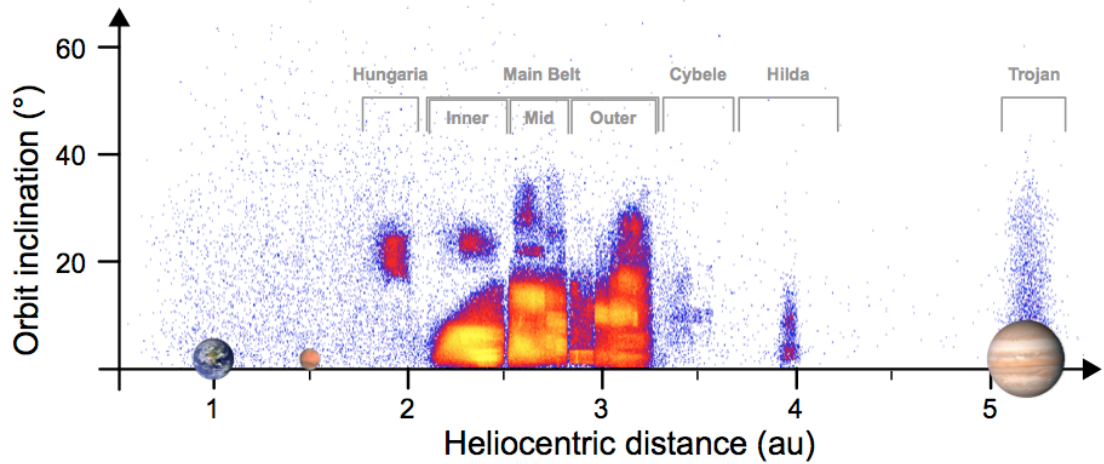

Fig. 1 The asteroid belt in context with the planets. This plot shows the location of the main belt with respect to the planets and the sun as well as the orbital structure of asteroid inclinations and number density of objects (yellow represents highest number density, blue the lowest). Asteroids have much higher orbital eccentricities and inclinations than do the planets. The structure of the belt is divided by unstable regions, seen most prominently at 2.5 and 2.8 au (locations where an asteroid's orbit is 'in resonance' with Jupiter's orbit), that separate the inner, middle, and outer sections of the main belt. The Hungaria asteroids are located closer to the Sun than is the main belt and have orbital inclinations centered near 20 degrees. The Hildas are located near 4 au and the Jupiter Trojans are in the L4 and L5 Lagrange points of Jupiter's orbit.

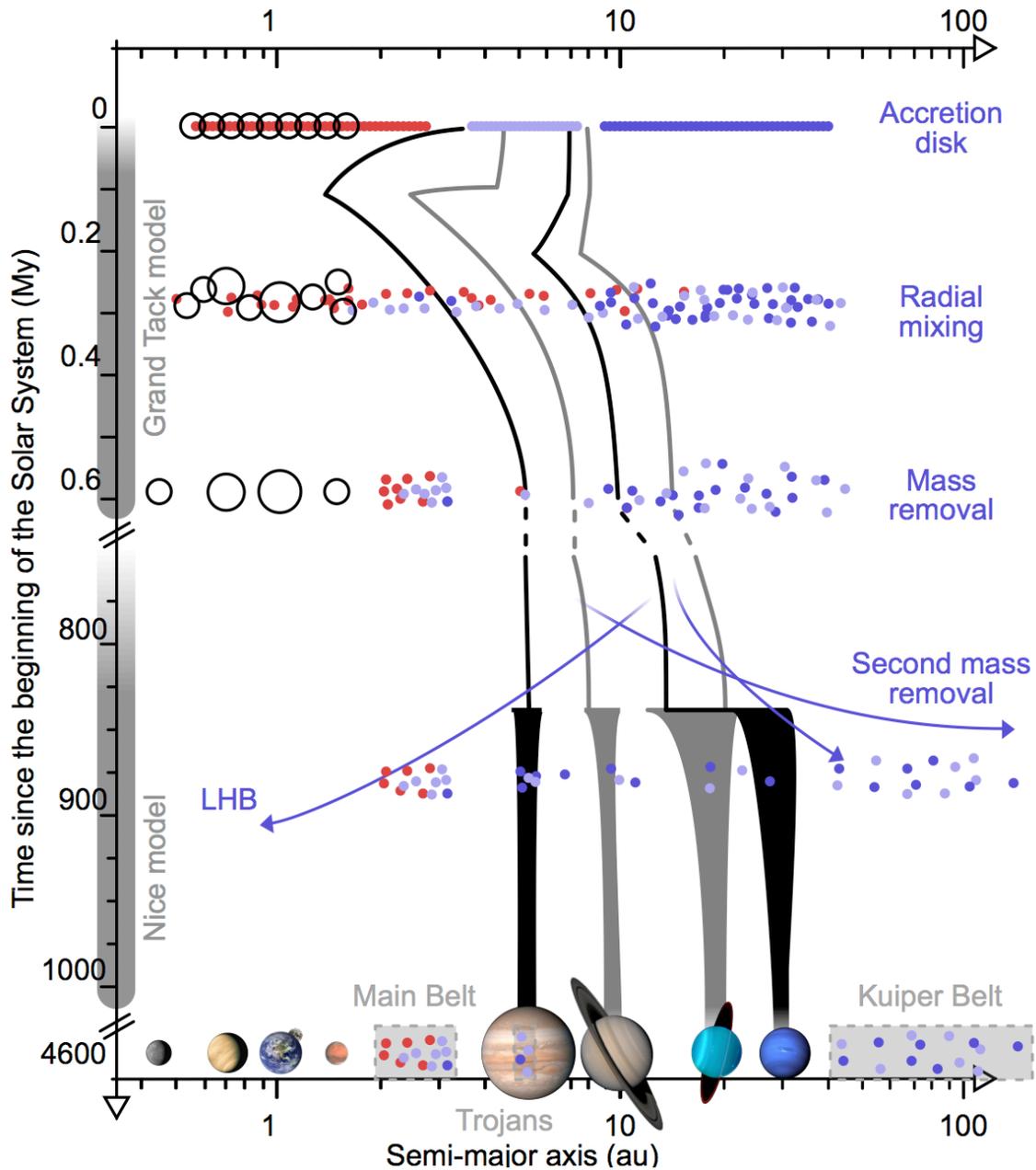

Fig 2. Cartoon of the effects of planetary migration on the asteroid belt. This figure captures some major components of the dynamical history of small bodies in the solar system based on models[11,12,50,53]. These models may not represent the actual history of the solar system, but are possible histories. They contain periods of radial mixing, mass removal, and planet migration-ultimately arriving at the current distribution of planets and small-body populations.

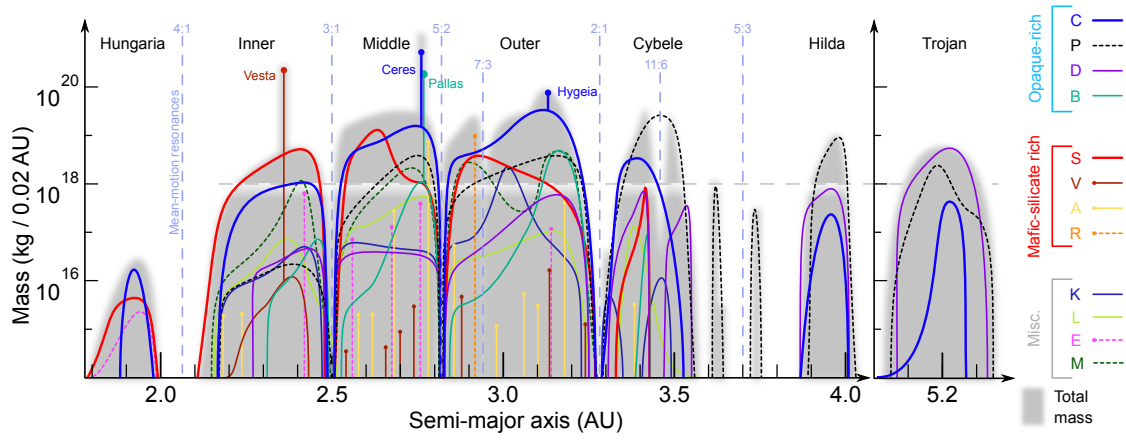

Fig 3. The compositional mass distribution throughout the asteroid belt out to the Trojans. The gray background is the total mass within each 0.02-au bin. Each color represents a unique spectral class of asteroid, denoted by a letter in the key. The horizontal line at $10^{18}$ kg is the limit of the work from the 1980s[2,8,9]. While the upper portion of the plot remains consistent with that work, immense detail is revealed at the lower mass range[19].

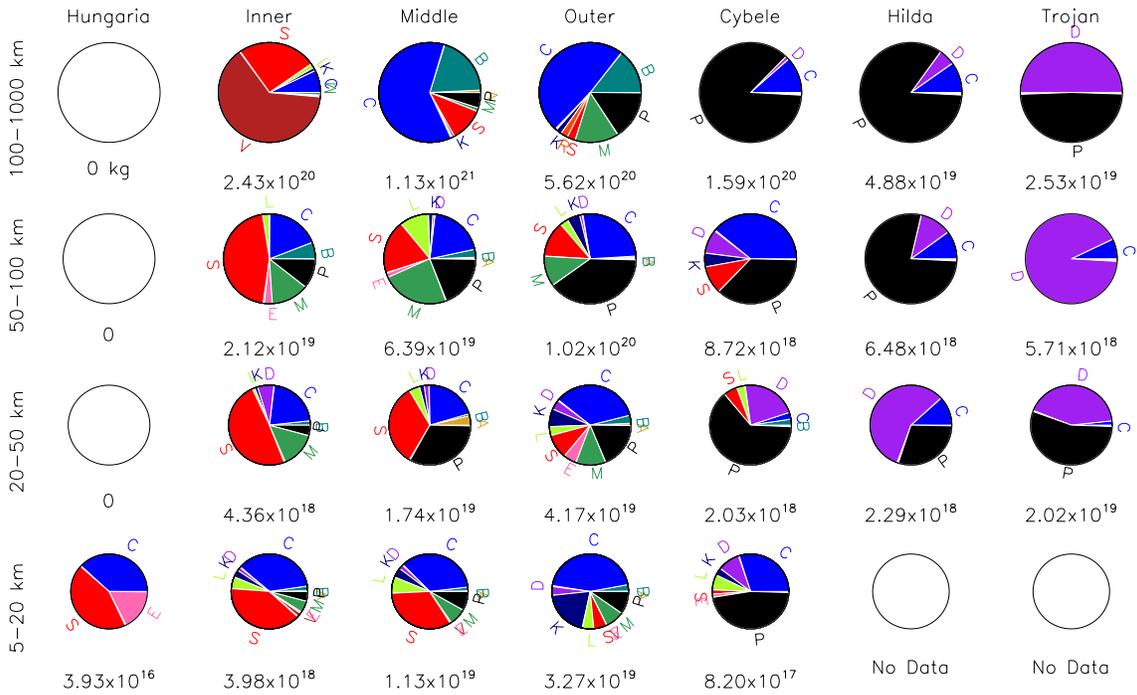

Fig 4. The compositional mass distribution as a function of size throughout the main belt out to the Trojans. The mass is calculated for each individual object with a diameter of 50km and greater using its albedo to determine size and the average density for that asteroids taxonomic class. For the smaller sizes we determine the fractional contribution of each class at each size and semi-major axis, and then apply that fraction to the distribution of all known asteroids from the Minor Planet Center including a correction for discovery incompleteness at the smallest sizes in the middle and outer belt[19]. Asteroid mass is grouped according to objects within four size ranges including diameters of 100-1000km, 50-100km, 20-50km, and 5-20km. Seven zones are defined as in Fig 1: Hungaria, Inner, Middle, and Outer Belt, Cybele, Hilda, and Trojan. The total mass of each zone at each size is labeled and the pie charts mark the fractional mass contribution of each unique spectral class of asteroid. The total mass of Hildas and Trojans are underestimated because of discovery incompleteness. The relative contribution of each class changes with both size and distance.

Box 1:

**The Hilda and Trojan Asteroids**

The Hilda and Jupiter Trojan asteroids are located beyond the main asteroid belt at 4au and 5.2au, respectively (Fig. 1). The asteroid types in these regions are physically distinct from the main belt and from each other: the largest Hildas are dominated by spectral P-type asteroids and the largest Trojans are dominated by D-types[2]. Despite interlopers of all types becoming more common throughout the main belt, the Hildas and Trojans remained curiously distinct and homogeneous.

Continued observations find that bright objects among Hildas and Trojans are scarce even at the small size scales[77-84]. It was recently and unexpectedly discovered, however, is the smallest bodies in these regions break rank (Fig. 4): most of the small Hildas have physical properties more similar to Trojans (D-type) and the makeup of the Trojans also changes with size with differing fractions of D-types and Hilda-like P-types[19,85-86]. Migration scenarios can now explain why the Hildas and Trojans looks so different from the main belt, but they cannot yet explain the important details of why they look distinct from each other at largest sizes (in relative fraction of the D-type and P-type asteroids) and are also different still at the smaller sizes.